# The temperature dependent geometric phase

Zheng-Chuan Wang

The University of Chinese Academy of Sciences, P. O. Box 4588, Beijing 100049, China, wangzc@ucas.ac.cn

## Abstract

There exists a geometric phase for a quantum state during the adiabatic evolution of the system. If the adiabatic procedure happens between the system and the environment interacting with it similar to Born-Oppenheimer (BO) approximation, we can introduce a temperature into the environment, which can be regarded as in an equilibrium state. Then a temperature-dependent geometric phase can be obtained for the system, which originates from the Abelian gauge potential induced by the BO approximation. This gauge potential contributes to the effective potential of the system, which is temperature dependent, too. Finally, we demonstrate them using an example of $H_2^+$ ion system.

# I. Introduction

Since proposed by Berry in 1984[1], the geometric phase has been widely explored in theories and experiments[2]. It was soon extended to the non-adiabatic case in 1986 by Aharonov et al.[3] and to the non-Abelian case in 1984 by Wilceck[4]. Except for the usual expression by the wavefunction of the Schrodinger equation, it had other expressions, i.e., the path integral[5], the relativistic Dirac equation[6] as well as the density matrix[7] et al.. As shown by Thouless et al., the geometric phase can play an important role in the adiabatic particle transport process[8], and it was also used to elaborate the quantized Hall conductance[9]. Till now, the geometric phase has been extensively investigated in atomic and molecular physics, optics, condensed matter physics and quantum field theory[10], in particular the topological character therein.

Usually, the study of geometric phase focus on the pure state of a quantum system, but how to extend it to the mixed state is still an open question. For a mixed state, which is compromised by many pure states with certain probability, there is no fixed relative phase between different pure states, so it is difficult to define the corresponding geometric phase between the pure states in a mixed state. In 1989, Uhlmann proposed a geometric phase for the mixed state and generalized the Berry curvature to the finite temperature case by means of the purification on the mixed state[11], where the geometric phase is defined in a pure state lying in an enlarged Hilbert space, so it depends on the auxiliary space for obtaining the Hilbert space enlarged. In 2019, Wang proposed a sub-geometric phase for the density matrix of the pure state[7], the sub-geometric phase appears in the off-diagonal elements of the matrix, but not in the diagonal elements. Since the density matrix of a mixed state can be expressed as $\rho = \sum_k p_k \rho_k$, where $\rho_k$ is the density matrix of the $k$-pure state in the mixed state, and $p_k$ is its probability appearing in the mixed state, the sub-geometric phase in the density matrix $\rho_k$ of a pure state will finally appear in the mixed state, too. If we further adopt $p_k$ as the Boltzmann distribution, then the sub-geometric phase can be extended to the finite temperature case. There are also other treatments to explore the geometric phase in mixed states[12,13].

It should be noted that in wang's treatment of the sub-geometric phase the temperature appears in the Boltzmann distribution, not in the sub-geometric phase of the pure state. Although the density matrix of a mixed state is temperature dependent, the sub-geometric phase doesn't depend on temperature at all. If we can find a geometric phase which is itself temperature dependent, then the geometric phase can be directly extended to the case of finite temperature, which is meaningful. In this manuscript, we will propose a temperature dependent geometric phase for a system interacting with its environment adiabatically, the temperature is introduced by the environment, which is assume to be in the equilibrium state.

# II. Theoretical formalism

Consider a quantum system interacting with its environment. The Hamiltonian for the system and environment can be written as follows:

$$H_S = \sum_i \frac{\vec{P}_i^2}{2M_i} + V_S(\{R_i\}) \tag{1}$$

and

$$H_E = \sum_i \frac{\vec{p}_i^2}{2m_i} + V_E(\{r_i\}) \tag{2}$$

respectively. If $V_{SE}(\{R_i\},\{r_i\})$ represents the interaction between the system and environment, then the total Hamiltonian for both the system and environment is

$$H = H_S + H_E + V_{SE}(\{R_i\},\{r_i\}), \tag{3}$$

where $\sum_i \frac{\vec{P}_i^2}{2M_i}$ and $\sum_i \frac{\vec{p}_i^2}{2m_i}$ are the kinetic energies of the system and environment, $V_S(\{R_i\})$ and $V_E(\{r_i\})$ are their potential energies, respectively. If the relaxation time in the environment is more faster than the system, the environment will arrive at the equilibrium state after a relaxation procedure, the variables $\{\vec{P}_i\}$ and $\{R_i\}$ of the system can be approximately regarded as the slow variables, whereas the variables $\{\vec{p}_i\}$ and $\{r_i\}$ for the environment can be regarded as fast variables. Similar to the Born-Oppenheimer approximation, we can solve the Schrodinger equation for the environment by fixing the slow variables $\{\vec{P}_i\}$ and $\{R_i\}$ of the system in the first, it is

$$[\sum_i \frac{\vec{p}_i^2}{2m_i} + V_E(\{r_i\})]\varphi(\{R_i\},\{r_i\}) = \varepsilon(\{R_i\})\varphi(\{R_i\},\{r_i\}). \tag{4}$$

Under the Hartree approximation, the above equation for the many particles of the environment can be simplified as an equation of a single particle in the Hartree potential $V_{Har}(r)$:

$$[\frac{\vec{p}^2}{2m} + V_{Har}(r)]\varphi(\{R_i\},r) = \varepsilon(\{R_i\})\varphi(\{R_i\},r). \tag{5}$$

If we write the Hartree wavefunction $\varphi(\{R_i\},r)$ as:

$$\varphi(\{R_i\},r) = A(\{R_i\},r)e^{\frac{i}{\hbar}S(\{R_i\},r)}, \tag{6}$$

where the amplitude $A(\{R_i\},r)$ and the phase (the action) $S(\{R_i\},r)$ corresponding to the wavefunction are all real functions. Substitute the above equation into Eq.(5), and we have

$$-\hbar^2\nabla^2 A(r) - 2i\hbar\nabla A(r)\nabla S(r) + A(r)\nabla S(r)\nabla S(r) - i\hbar A(r)\nabla^2 S(r) = (\varepsilon(\{R_i\}) - V_{Har}(r))A(r) \tag{7}$$

If we further expand the action function $S(r)$ in powers of $\hbar$,

$$S(r) = S_0(r) + \frac{\hbar}{i}S_1 + (\frac{\hbar}{i})^2 S_2 + ..., \tag{8}$$

and substitute it into Eq.(7), the zero order of $\hbar$ gives

$$A(r)\nabla S_0(r)\nabla S_0(r) = (\varepsilon(\{R_i\}) - V_{Har}(r))A(r), \tag{9}$$

while the first order of $\hbar$ gives

$$2\nabla A(r)\nabla S_0(r) + 2A(r)\nabla S_0(r)\nabla S_1(r) - A(r)\nabla^2 S_0(r) = 0, \tag{10}$$

and the second order of $\hbar$ gives

$$-\nabla^2 A(r) - 2\nabla A(r)\nabla S_1(r) - 2A(r)\nabla S_0(r)\nabla S_2(r) - A(r)\nabla S_1(r)\nabla S_1(r) - A(r)\nabla^2 S_1(r) = 0. \tag{11}$$

From Eq.(9), we can obtain the zero order of action as follows:

$$S_0(r) = \int \sqrt{\varepsilon(\{R_i\}) - V_{Har}(r)} dr. \qquad (12)$$

Since the relaxation time of particles in the environment is much faster than the relaxation time of the system, the environment can rapidly arrive at the equilibrium state and finally remain in equilibrium. The probability density of particles in the environment is then proportional to the Boltzmann distribution,

$$|\varphi(r)|^2 = n_0 e^{-\frac{\varepsilon}{k_B T(r)}}, \qquad (13)$$

where $T(r)$ is the temperature in the environment, which can be regarded as in the local equilibrium, $k_B$ is the Boltzmann constant. The constant $n_0$ can be determined by the normalizing condition of $\varphi(r)$. So we can obtain the amplitude $A(r)$ of wavefunction $\varphi(r)$ as follows:

$$A(r) = \sqrt{n_0 e^{-\frac{\varepsilon}{k_B T(r)}}}, \qquad (14)$$

substitute it into Eq.(10), the first order of action can be obtained as

$$S_1(r) = \int \frac{-A(r)\nabla^2 S_0(r) - 2\nabla A(r)\nabla S_0(r)}{2A(r)\nabla S_0(r)} dr. \qquad (15)$$

Order by order, we can obtain the higher order actions step by step.

Since the variables $\{\vec{P}_i\}$ and $\{R_i\}$ for the system can be regarded as slow variables, while the variables $\{\vec{p}_i\}$ and $\{r_i\}$ for the environment can be fast variables, if the motion of the system is adiabatic, similar to the BO approximation adopted in molecular physics, using Eq.(5), under the Hartree approximation, we can obtain the motion equation for a single particle in the system as

$$[\frac{1}{2M}(\vec{P} - \hbar\vec{A})^2 + V_{eff}(R)]\psi(R) = E\psi(R), \qquad (16)$$

where $\vec{A} = i < \varphi(R,r)|\nabla_R|\varphi(R,r) >$ is the gauge potential induced by the adiabatic variation of the coordinate $R$ of the particles in the system, which brings a geometric phase accompanying with the wavefunction $\psi(R)$ as:

$$\gamma = \oint \vec{A} \cdot d\vec{R}. \qquad (17)$$

According to Eq.(14), the amplitude $A(r)$ of wavefunction $\varphi(R,r)$ is temperature dependent. The gauge potential and geometric phase are temperature dependent, which is the central result in this manuscript. The $V_{eff}(R)$ is the effective potential for the motion of particles in the system, it can be expressed as follows:

$$V_{eff}(R) = \varepsilon(R) + \frac{\hbar^2}{2M}(<\nabla_R\varphi(R,r)|\nabla_R\varphi(R,r)> - \vec{A}^2), \qquad (18)$$

which indicates that the effective potential is temperature dependent, too.

In the adiabatic approximation, we can write the total wavefunction for the whole system and environment as $|\Psi(R,r)> = |\psi(R)\varphi(r)>$, and its corresponding density matrix operator is $\hat{\rho}(R,r) = |\Psi(R,r)><\Psi(R,r)|$, which is also temperature dependent. Thus the physical observable corresponding to the operator $\hat{B}$ can be expressed as follows:

$$\bar{B} = Tr(\hat{\rho}(R,r)\hat{B}), \qquad (19)$$

which is temperature dependent, where $\hat{\rho}(R,r)$ is not the density matrix in equilibrium, it is a

density matrix in nonequilibrium, which satisfies the following Liouville equation

$$\frac{d\hat{\rho}(R,r,t)}{dt} = \frac{1}{i\hbar}[\hat{H}, \hat{\rho}(R,r,t)]. \tag{20}$$

Eq.(19) can be seen as the thermal statistical average for operator $\hat{B}$, which is different from the statistical average in equilibrium.

### III. Example

For simplicity, we take $H_2^+$ as an example to investigate the temperature-dependent gauge potential and geometric phase. In atomic units, the Hamiltonian for $H_2^+$ can be written as:

$$H = H_e + \frac{1}{R}, \tag{21}$$

where $H_e = -\frac{1}{2}\nabla^2 - \frac{1}{r_a} - \frac{1}{r_b}$ is the Hamiltonian for the electron with $r_a$ and $r_b$ the distance between the electron and the two nuclei, $R$ is the distance between two nuclei. The electronic wavefunction for the ground state is

$$\psi_\pm = C_\pm(\psi_a \pm \psi_b), \tag{22}$$

where $C_\pm = (2 \pm 2Y)$ is the normalizing constant, where $Y = (\psi_a, \psi_b)$, $\pm$ correspond to the even and odd parity states, respectively. According to Eq.(13), the wavefunction $\psi_a$ and $\psi_b$ can be approximated as the temperature-dependent form, they are:

$$\psi_a = \frac{\lambda^{3/2}}{\sqrt{\pi}} e^{-\lambda r_a} \text{ and } \psi_b = \frac{\lambda^{3/2}}{\sqrt{\pi}} e^{-\lambda r_b}, \tag{23}$$

where $\lambda = \left(\pi n_0 e^{-\frac{\varepsilon}{k_B T(r)}}\right)^{1/3}$. The energies corresponding to $\psi_\pm$ are

$$E_\pm = \frac{1}{R} - \frac{1}{2}\lambda^2 + \frac{\lambda(\lambda-1) - \Pi \pm (\lambda-2)\Sigma}{1 \pm Y}, \tag{24}$$

where $\Pi = \frac{1}{R}[1 - (1 + \lambda R)e^{2\lambda R}]$, $\Sigma = \lambda(1 + \lambda R)e^{-\lambda R}$.

Under the adiabatic approximation, the gauge potential for the even parity is $\vec{A}_+ = i < \psi_+(R)|\nabla_R|\psi_+(R) >$, and for the odd parity is $\vec{A}_- = i < \psi_-(R)|\nabla_R|\psi_-(R) >$, their corresponding geometric phases are $\theta_+ = i \int_0^{R_m} < \psi_+(R)|\nabla_R|\psi_+(R) > dR$ and $\theta_- = i \int_0^{R_m} < \psi_-(R)|\nabla_R|\psi_-(R) > dR$, respectively. Both the gauge potential and geometric phase are temperature dependent. Since the energy $E_-$ corresponding to the odd parity state $|\psi_-(R) >$ has no minimum, it can't constitute the stable structure for $H_2^+$. In the next we only study the even parity state $|\psi_+(R) >$.

In Fig.1, we plot the gauge potential $\vec{A}_+$ corresponding to $|\psi_+(R) >$ as a function of $R$ at different temperatures of 100K, 200K and 300K. We can see that the higher the temperature, the smaller the maximum for the gauge potential, all the three curves tend to zero when the distance $R$ increases, where we adopted the atomic units for the distance $R$ between two nuclei. The geometric phase vs temperature is shown in Fig.2, where $R_m$ in the integral $\theta_+ = i \int_0^{R_m} < \psi_+(R)|\nabla_R|\psi_+(R) >$ is adopted as 5 atomic units. Obviously the geometric phase is temperature dependent, too. It nearly decreases with temperature except at the low temperature around 100K. In Fig.3, we plot the effective potential for the motion of the nucleus as a function of distance $R$, it is shown that the $H_2^+$ will become stable at the minimum of the effective potential. The minimum points of the effective potentials are 2.10, 2.12 and 2.15 (atomic units), respectively, which is comparable to the experimental value of 2.004 (atomic units), indicating that the temperature can change the stable distance of two nuclei slightly.

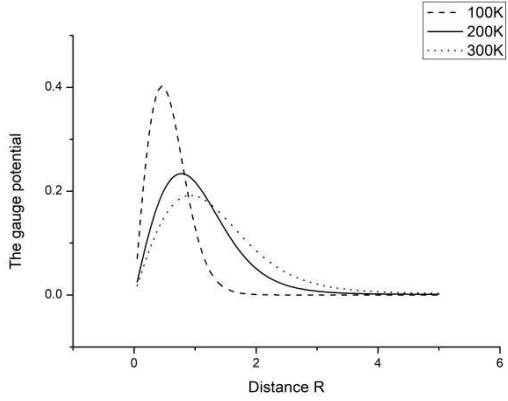

Fig.1 The gauge potential vs. distance of nuclei at different temperatures of 100K, 200K and 300K, where we adopted the atomic units.

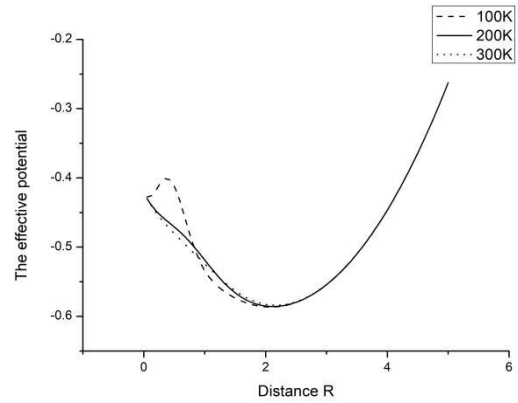

Fig.3 The effective potential vs. distance of nuclei at different temperatures of 100K, 200K and 300K, where we adopted the atomic units.

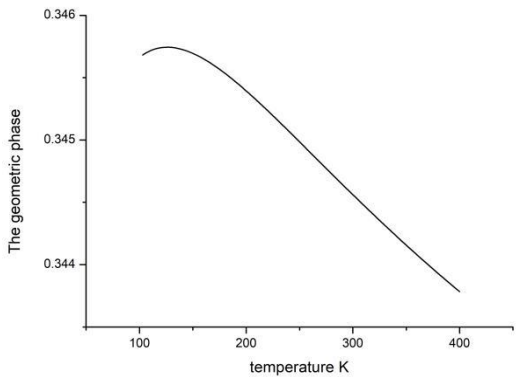

Fig.2 The geometric phase vs. temperature, where we adopted the atomic units.

## IV. Summary and discussions

In this manuscript, we propose the temperature-dependent gauge potential and geometric phase in a system interacting with its environment adiabatically. The temperature can be introduced by the wavefunction of particles in the environment which can be regarded as remaining in the equilibrium state, because we assume that the relaxation procedure of the environment is more faster than the particles in the system. When we adopt the adiabatic approximation for the evolution of the system and environment, a gauge potential will be induced similar to the Born-Oppenheimer approximation in molecular physics, which is temperature dependent, as stated. Certainly, the effective potential for the motion of particles in the system is also temperature dependent, because it contains the temperature dependent gauge potential which contributes a geometric phase to the wavefunction of particles in the system. In Fig.1 and Fig.2, we plot the gauge potential and geometric phase corresponding to the even parity state, respectively, the effect of temperature-dependence is obvious.

It should be noted that we adopted the adiabatic approximation in the derivation of temperature dependent gauge potential, and assumed that the particles in the environment remain in the equilibrium state. If the system doesn't follow the adiabatic evolution, we can not obtain the gauge potential similar to the Born-Oppenheimer approximation, and if the articles in the environment are not in equilibrium, we can not introduce the temperature into the gauge potential and geometric phase. How to extend the temperature geometric phase to the case of non-adiabatic and non-equilibrium is still an open question, we leave it for further exploration in the future.

## Acknowledgments

This study is supported by the National Key R&D Program of China (Grant No. 2022YFA1402703).

## Data Availability Statement

Data sets generated during the current study are available from the corresponding author on reasonable request.

## Additional information

Competing interest statement: The authors declare that they have no competing interests.